\documentclass[10pt, final, conference, letterpaper, onecolumn, oneside]{./IEEEtran}

\usepackage{cite}
\usepackage{graphicx}
\usepackage{subfigure}
\usepackage{amsmath}
\usepackage{amssymb}
\usepackage{mathtools}
\usepackage{bm}
\usepackage{epsfig}
\usepackage{times}
\usepackage{color}
\usepackage{url}
\usepackage{array}
\usepackage{multirow}
\usepackage{rotating}
\usepackage{extarrows}
\usepackage{mathabx} 
\usepackage{wasysym}
\usepackage{./clrscode_mod}

\newtheorem{pavikl}{\textbf{Lemma}}

\newcommand{\argmin}{\operatornamewithlimits{argmin}}
\newcommand{\argmax}{\operatornamewithlimits{argmax}}

\newcommand{\halfrac}{\frac{1}{2}}

\begin{document}

\title{Analog Network Coding in Nonlinear Chains}%
\author{\IEEEauthorblockN{Samar Agnihotri}%
\IEEEauthorblockA{School of Computing and Electrical Engineering, Indian Institute of Technology Mandi, HP - 175$\,$001, India}%
Email: samar.agnihotri@gmail.com%
}

\maketitle

\begin{abstract}
The problem of characterizing the optimal rate achievable with analog network coding (ANC) for a unicast communication over general wireless relay networks is computationally hard. A relay node performing ANC scales and forwards its input signals. The source-destination channel in such communication scenarios is, in general, an intersymbol interference (ISI) channel which leads to the single-letter characterization of the optimal rate in terms of an optimization problem with nonconvex, non closed-form objective function and non-convex constraints. For a special class of such networks, called layered networks, a few key results and insights are however available.

To gain insights into the nature of the optimal solution and to construct low-complexity schemes to characterize the optimal rate for general wireless relay networks, we need (1) network topologies that are regular enough to be amenable for analysis, yet general enough to capture essential characteristics of general wireless relay networks, and (2) schemes to approximate the objective function in closed-form without significantly compromising the performance. Towards these two goals, this work proposes (1) nonlinear chain networks, and (2) two approximation schemes. We show that their combination allows us to tightly characterize the optimal ANC rate with low computational complexity for a much larger class of general wireless relay networks than possible with existing schemes.
\end{abstract}

\section{Introduction}
\label{sec:intro}
In a wireless network, signals transmitted simultaneously by multiple sources add in the air. Each node receives a \textit{noisy sum} of these signals, \textit{i.e.} a linear combination of the received signals and noise. A multihop relay scheme where an intermediate relay node merely amplifies and forwards this noisy sum is referred to as Analog Network Coding (ANC) \cite{107kattiGollakottaKatabi, 110maricGoldsmithMedard}. Therefore, ANC extends to multihop wireless networks the idea of linear network coding \cite{103liYeungCai} where an intermediate node sends out a linear combination of its incoming packets. 

The performance of the analog network coding is considered in \cite{110maricGoldsmithMedard, 111liuCai} in a special class of relay networks called layered relay networks in the system setting of real channel gains, and full-duplex relays operating under maximum average power constraint. Further, the achievable rate is computed under two assumptions: (A) each relay node scales the received signal to the maximum extent possible subject to its transmit power constraint, (B) the nodes in all but at most one layer operate in the high-SNR regime. It is shown that the rate achieved under these two assumptions approaches network capacity as the source power increases. In a series of papers \cite{112agnihotriJaggiChen, 112agnihotriJaggiChen2, 112agnihotriJaggiChen3}, Agnihotri \textit{et al.} consider the problem of the optimal ANC rate in general layered networks, but without the aforementioned two assumptions on relay operation. The results therein offer a few key insights into design and operation of low-complexity schemes to compute the optimal ANC rate in such networks. Specifically, \cite{112agnihotriJaggiChen} shows that in layered networks, the scaling factors for all relay nodes that lead to the optimal ANC rate at the destination can be computed in a layer-by-layer manner. Further, \cite{112agnihotriJaggiChen2} establishes that in the layered networks with only a single layer of relay nodes, the scaling factors that achieve the optimal ANC rate at the destination can be computed in a greedy manner, thus providing an alternate approach to the results in \cite{109jingJafarkhani}. Finally, \cite{112agnihotriJaggiChen3} shows that in the layered networks deploying only $k$ out of $n$ relays in each layer, $k < n, 1 < n$, leads to significant reduction in the computational complexity of the optimal ANC rate computation with only logarithmic (in $n$ and $k$) loss in the optimal rate.

Further, in \cite{111agnihotriJaggiChen} single-letter characterization of the optimal ANC rate in general relay networks is obtained in the aforementioned system setting. However, the characterization proposed therein is in terms of an optimization problem which is computationally hard to solve exactly for all but some trivial scenarios.

We argue that it is important to obtain tight characterization of the optimal ANC rate achievable in general wireless relay networks with low-complexity schemes for three reasons. First, an ANC scheme based on a relaying scheme as simple as amplify-and-forward allows us to benchmark the performance of other physical-layer relaying schemes, such as \textit{Estimate-and-Forward} \cite{107gomadamJafar}. Second, such a scheme helps us obtain insights on the information flow and optimum relay operation in multihop relay networks in various SNR regimes. Finally, such insights may lead to construction of better physical-layer network coding schemes.

In order to construct low-complexity schemes to tightly approximate the optimal ANC rate in general relay networks, in this paper we propose a two pronged approach. First, we envision that by considering non-layered relay networks as a generalization of layered relay networks, result and insight in \cite{112agnihotriJaggiChen, 112agnihotriJaggiChen2, 112agnihotriJaggiChen3} can be used to obtain tight and computationally efficient approximation of the optimal ANC rate in general non-layered networks. However, as no general procedure exists to transform a given layered network into a given non-layered network and vice-versa without performance loss, results and insights for layered networks cannot be carried over to non-layered networks in some straightforward manner. Therefore, we propose to consider intermediate network topologies that are regular enough to allow straightforward correspondence with layered networks, yet general enough to capture essential characteristics of general wireless relay networks. Second, the objective function in the optimization problem formulation of the optimal ANC rate for general relay networks in \cite{111agnihotriJaggiChen} is in non-closed form, making it analytically intractable. Therefore, we need schemes to approximate the objective function in closed-form without significantly compromising the performance.

Our main contribution in this paper is twofold. First, we introduce nonlinear chain networks. In the layered networks, the relay nodes between a given source-destination pair are arranged in $L$ layers such that the nodes within a layer do not communicate among themselves but communicate only with the nodes in the next layer. This results in all paths from the source to the destination to be of the same length. Assuming identical delays along all links and relay nodes, this results in the signal received at the destination to be free from \textit{intersymbol interference (ISI)}. In general non-layered networks, on the other hand, a relay node can communicate with any subset of nodes in the network. This results in the source signal reaching the destination via multiple paths of possibly different lengths, causing ISI at the destination among the signals received along multiple paths. The nonlinear chain networks are positioned between these two extremes. In the nonlinear chain networks, the relay nodes between a given source-destination pair are arranged in a linear chain, but each node can communicate with $k, 2 \le k \le L$, forward nodes. Thus, the nonlinear chain networks can be considered as layered networks with only a single node in each layer. However, unlike the nodes in layered networks, the node in any layer in nonlinear chain communicates with the nodes in forward $k$ layers. In future, we plan to generalize nonlinear chain networks to more powerful intermediate network topologies.

Second, we introduce two schemes to approximate and tightly bound the optimal ANC rate with low computational complexity. These schemes allow us to tightly characterize the maximum ANC rate in a wider class of non-layered networks with polynomial-time complexity that cannot be so addressed using existing approaches.

\textit{Organization:} In Section~\ref{sec:sysModel}, we introduce nonlinear chain networks as a generalization of layered networks, and formulate the problem of maximum ANC rate in such networks. Section~\ref{sec:approxSchemes} introduces two schemes to approximate and bound the optimal ANC rate. Then in Section~\ref{sec:cmplxity_linearChains} we illustrate application of the two schemes to a small nonlinear chain network and characterize a few classes of nonlinear chain networks where the optimal rate can be tightly approximated with polynomial-time complexity. In Section~\ref{sec:perfAnalysis} we evaluate performance of the two schemes in approximating the optimal ANC rate for larger nonlinear chain networks in different SNR regimes. Finally, Section~\ref{sec:conclFW} concludes the paper.

\section{System Model}
\label{sec:sysModel}
\begin{figure}[!t]
\centering
\includegraphics[width=3.5in]{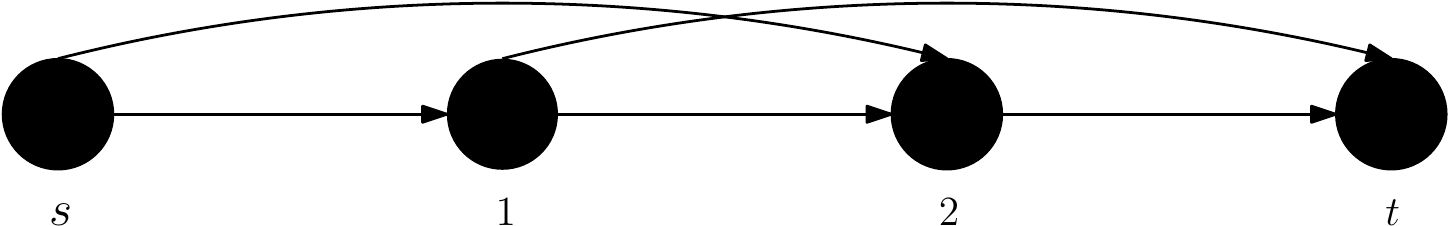}
\caption{A single source-single destination communication channel over $(2,2)$ nonlinear chain with two relay nodes between the source $s$ and the destination $t$ and each node communicating with two forward nodes.}
\label{fig:linChainExa}
\end{figure}

Consider a $(N+2)$-layer wireless relay network with directed links. The source $s$ is at layer `0', the destination $t$ is at layer `$N+1$', and a set $R$ of $N$ relay nodes is arranged in $N$ layers between them. Each node, except the destination $t$, communicates with $k$ forward nodes, $1 \le k \le N+1$. We call such networks as $(N,k)$ nonlinear chains. An instance of such a network is given in Figure~\ref{fig:linChainExa}. Every node is assumed to have a single antenna and operate in full-duplex mode.

\textit{Remark 1:} The classical three-terminal relay channel of \cite{071meulen} can be considered as $(1,2)$ nonlinear chain network.

At instant $n$, the channel output at node $i, i \in R \cup \{t\}$, is
\begin{equation}
\label{eqn:channelOut}
y_i[n] = \sum_{j \in {\mathcal N}(i)} h_{ji} x_j[n] + z_i[n], \quad - \infty < n < \infty,
\end{equation}
where $x_j[n]$ is the channel input of the node $j$ in the neighbor set ${\mathcal N}(i)$ of node $i$ with $|{\mathcal N}(i)| = \min(k, N+1-i)$. In \eqref{eqn:channelOut}, $h_{ji}$ is a real number representing the channel gain along the link from the node $j$ to the node $i$. It is assumed to be fixed (for example, as in a single realization of a fading process) and known throughout the network. The noise process $\{z_i[n]\}$ is a sequence (in $n$) of \textit{i.i.d.} Gaussian random variables with $z_i[n] \sim {\cal N}(0, \sigma^2)$. We also assume that $z_i$ are independent of the input signal and of each other. The source symbols $x_s[n], - \infty < n < \infty$, are \textit{i.i.d.} Gaussian random variables with zero mean and variance $P_s$ that satisfy an average source power constraint, $x_s[n] \sim {\cal N}(0, P_s)$. We assume that the $i^{\textrm{th}}$ relay's transmit power is constrained as:
\begin{equation}
\label{eqn:pwrConstraint}
E[x_i^2[n]] \le P_i, \quad - \infty < n < \infty
\end{equation}

In analog network coding each relay node amplifies and forwards the noisy signal sum received at its input. More precisely, a relay node $i$ at instant $n+1$ transmits the scaled version of $y_i[n]$, its input at time instant $n$, as follows
\begin{equation}
\label{eqn:AFdef}
x_i[n+1] = \beta_i y_i[n], \quad 0 \le \beta_i^2 \le \beta_{i,max}^2 = P_i/P_{R,i},
\end{equation}
where $P_{R,i}$ is the received power at the node $i$.

The input-output channel between the source and the destination for $(N,k)$ chain network is an intersymbol interference (ISI) channel that at instant $n$ is given by (using \eqref{eqn:channelOut} and \eqref{eqn:AFdef}):
\begin{align}
y_t[n] = &\sum\limits_{d=\lfloor\frac{N}{k}\rfloor}^N\left(\sum\limits_{(i_1,...,i_d) \in K_d}h_{s i_1}\beta_{i_1}h_{i_1 i_2}...h_{i_{d-1} i_d}\beta_{i_d} h_{i_d t}\right)x_s[n-d] \label{eqn:ISIchnl} \\
         &+ \sum\limits_{m=1}^N\sum\limits_{d=\lfloor\frac{N-m}{k}\rfloor+1}^{N-m+1}\left(\sum\limits_{\stackrel{(i_1,...i_d)}{\in K_{m,d}}}\beta_{m}h_{m i_1}...h_{i_{d-1} i_d}\beta_{i_d}h_{i_d t}\right)z_m[n-d] + z_t[n] \nonumber
\end{align}
where $K_d$ is the set of $d$-tuples of node indices corresponding to all the paths from the source to the destination with delay $d, \lfloor\frac{N}{k}\rfloor \le d \le N$. Note that the length of the longest such path is $N$ and the length of the shortest path is $\lfloor\frac{N}{k}\rfloor$. Similarly, $K_{m,d}$ is the set of $d$-tuples of node indices corresponding to all paths from the $m^{th}$ relay to the destination with path delay $d$, $\lfloor\frac{N-m}{k}\rfloor+1 \le d \le D^m = (N-m+1)$, $1 \le m \le N$. The following lemma computes the number of such paths.

\begin{pavikl}
\label{lemma:noOpaths}
In a $(N,k)$ nonlinear chain, number of paths from the source $s$ to the destination $t$ with delay $d$ is:
\begin{equation*}
|K_d| = \sum_{r=0}^{d+1} (-1)^r C(d+1, r)\,C(N-rk, d)
\end{equation*}
Similarly, number of paths from the $m^{th}$ relay to the destination $t$ with delay $d$ is:
\begin{equation*}
|K_{m,d}| = \sum\limits_{r=0}^{d} (-1)^r C(d, r)\,C(N-m-rk, d-1)
\end{equation*} 
\end{pavikl}

Following \cite{111agnihotriJaggiChen}, introduce \textit{modified} channel gains as follows. For all the paths between the source $s$ and the destination $t$:
\begin{equation}
\label{eqn:modChnlParams}
h_d = \sum\limits_{\stackrel{(i_1,...,i_d)}{\in K_d}}h_{s i_1}\beta_{i_1}h_{i_1 i_2}...h_{i_{d-1} i_d}\beta_{i_d} h_{i_d t}, \lfloor N/k \rfloor \le d \le N
\end{equation}
For all the paths between the $m^{\textrm{th}}$ relay, $1 \le m \le N$, and $t$:
\begin{align}
& h_{m,0} = 0, \label{eqn:modChnlParams2}\\
& h_{m,d} = \sum\limits_{\stackrel{(i_1,...i_d)}{\in K_{m,d}}}\beta_{m}h_{m i_1}...h_{i_{d-1} i_d}\beta_{i_d}h_{i_d t}, \left\lfloor \frac{N-m}{k} \right\rfloor+1 \le d \le D^m \nonumber
\end{align}

In terms of these modified channel gains the source-destination ISI channel in \eqref{eqn:ISIchnl} can be written as:
\begin{equation}
\label{eqn:ISIchnlmod}
y_t[n]=\sum\limits_{d=\lfloor \frac{N}{k}\rfloor}^N  h_d x_s[n-d] + \sum\limits_{m=1}^N \sum\limits_{d=\lfloor \frac{N-m}{k}\rfloor+1}^{N-m+1}  h_{m,d} z_m[n-d]+z_t[n]
\end{equation}

\textit{Problem Formulation:} For a given network-wide scaling vector $\bm{\beta} = (\beta_1, \ldots,\beta_N)$, the achievable rate for the channel in \eqref{eqn:ISIchnlmod} with \textit{i.i.d.} Gaussian input is \cite[Lemma 1]{111agnihotriJaggiChen}:
\begin{equation}
\label{eqn:infoRateFin}
I(P_s, \bm{\beta}) = \frac{1}{2 \pi} \int_{0}^{\pi} \log\bigg[1 + \frac{P_s}{\sigma^2} \frac{|H(\lambda)|^2}{1 + \sum_{m=1}^N |H_{m}(\lambda)|^2}\bigg] d\lambda, 
\end{equation}
where with $i=\sqrt{-1}$
\begin{equation}
\label{eqn:trnsFn}
H(\lambda)=\sum\limits_{d=\lfloor\frac{N}{k}\rfloor}^N h_d e^{-id\lambda}, \: H_m(\lambda)=\sum\limits_{d=\lfloor\frac{N-m}{k}\rfloor+1}^{N-m+1} h_{m,d} e^{-id\lambda} 
\end{equation}
The maximum information-rate $I_{ANC}(P_s)$ achievable in a given nonlinear chain network with \textit{i.i.d.} Gaussian input is defined as the maximum of $I(P_s, \bm{\beta})$ over all feasible $\bm{\beta}$, subject to per relay transmit power constraint \eqref{eqn:AFdef}. That is:
\begin{equation}
\label{eqn:maxAFrate}
I_{ANC}(P_s) \stackrel{def}{=} \max_{\stackrel{\bm{\beta}:0 \le \beta_{m}^2 \le \beta_{m, max}^2}{1 \le m \le N}} I(P_s, \bm{\beta})
\end{equation}
This problem is computationally-hard for all but some trivial network instances and unrealistic assumptions on relay operation \cite{111agnihotriJaggiChen}. However, a closer analysis of nature of the integrand in \eqref{eqn:infoRateFin} allows us to express the objective function is closed-form which further allows us to construct low-complexity schemes to closely approximate and tightly bound the solution of \eqref{eqn:maxAFrate} for a wider class of networks than hitherto possible. In the next section, we discuss two such schemes.

\section{Two Approximation Schemes}
\label{sec:approxSchemes}
Let the integrand in \eqref{eqn:infoRateFin} is denoted as $g(\lambda)$. Then it can be rewritten as:
\begin{align*}
g(\lambda) &\stackrel{(a)}{=} \log\left[1 + \frac{P_s}{\sigma^2} \frac{\sum\limits_{i=0}^{N-\lfloor \frac{N}{k}\rfloor}A'_i \cos^i\lambda}{\sum\limits_{i=0}^{N-\lfloor \frac{N-1}{k}\rfloor-1}B'_i \cos^i \lambda}\right] \\
           &\stackrel{(b)}{=} \frac{1}{\sqrt{1-u^2}} \log\left[1+\frac{P_s}{\sigma^2}\frac{\sum\limits_{i=0}^{N-\lfloor \frac{N}{k}\rfloor}A'_i u^i}{\sum\limits_{i=0}^{N-\lfloor \frac{N-1}{k}\rfloor-1}B'_i u^i}\right] \\
           &= \frac{1}{\sqrt{1-u^2}} \log(1+f(u)),
\end{align*}
where $(a)$ follows from substituting the transfer functions in \eqref{eqn:trnsFn} into \eqref{eqn:infoRateFin} (for the details, please refer to Appendix B of the longer version of \cite{111agnihotriJaggiChen}) and $(b)$ follows from the substitution $u = \cos \lambda$. Thus \eqref{eqn:infoRateFin} can be rewritten as:
\begin{equation}
\label{eqn:infoRateFin2}
I(P_s, \bm{\beta}) = \frac{1}{2 \pi} \int_{-1}^{1} \frac{1}{\sqrt{1-u^2}} \log(1+f(u)) du
\end{equation}

\begin{figure}[!t]
\centering
\includegraphics[width=3.5in]{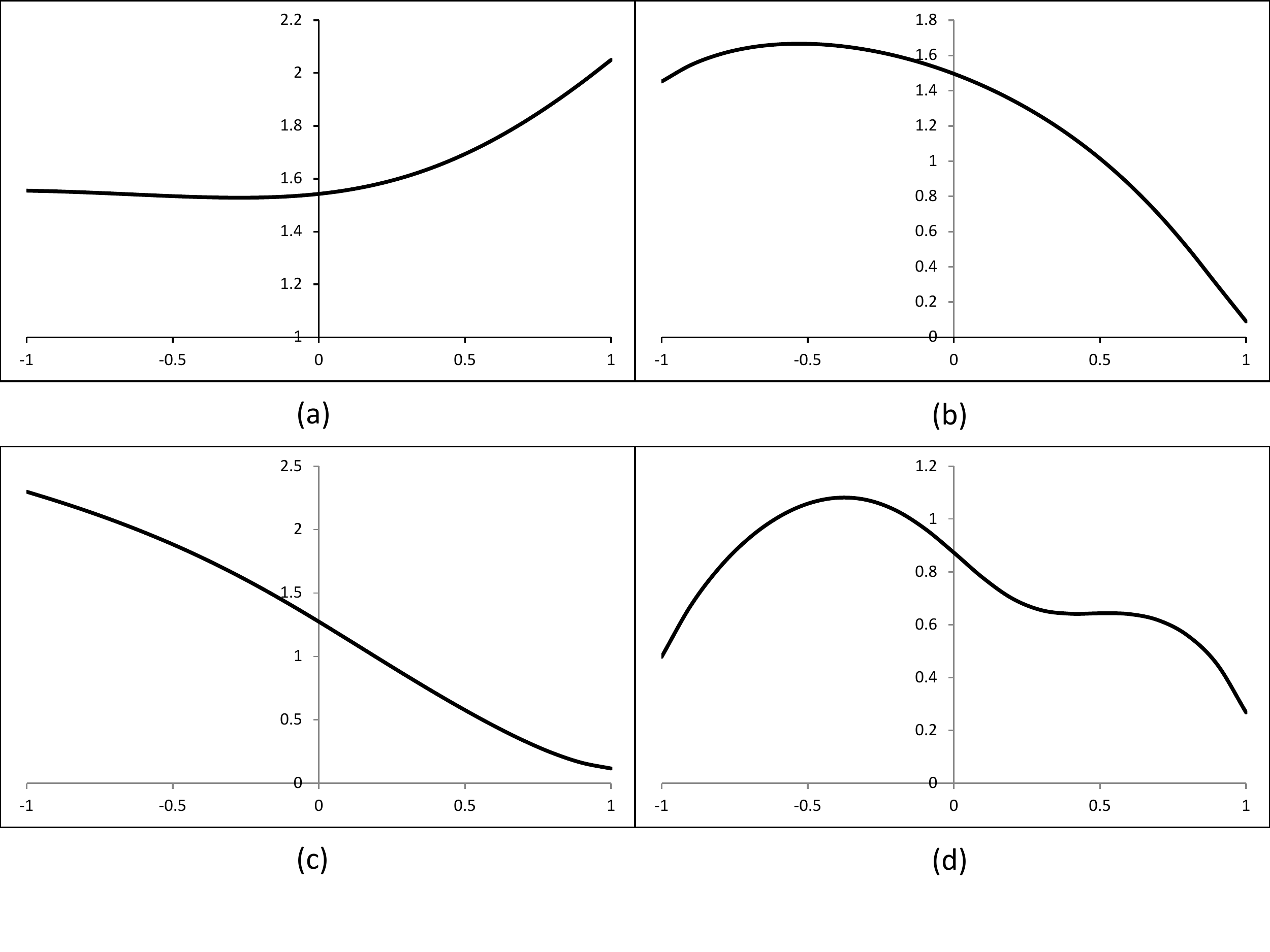}
\caption{Typical behavior of $\log(1+f(u))$ in range $[-1,1]$: (a) convex, (b) concave, (c) quasiconcave, and (d) irregular.}
\label{fig:fooPlots}
\end{figure}

\begin{pavikl}
\label{lemma:foo}
$f(u) > -1$, $u \in [-1, 1]$.
\end{pavikl}
This lemma implies that the function $\log(1+f(u))$ is bounded over the range of integration in \eqref{eqn:infoRateFin2}. Typical behavior of $\log(1+f(u))$ in range $[-1,1]$ is depicted in Figure~\ref{fig:fooPlots}. This implies that over the range of integration:
\begin{itemize}
\item \textit{Zeroth-order Approximation:} $\log(1+f(u))$ can be approximated by a straight line passing through $[\log(1+f(-1)),-1]$ and $[\log(1+f(1)),1]$ with slope $s = \halfrac\left[\log\frac{1+f(1)}{1+f(-1)}\right]$
\item \textit{First-order Approximation:} $\log(1+f(u))$ can be bounded between two tangent lines with slope equal to $s$ with one of the tangents passing through $[\log(1+f(u_{max})),u_{max}]$ and other through $[\log(1+f(u_{min})),u_{min}]$, where $u_{max}$ and $u_{min}$ are the values of the integration variable $u$ at which $\log(1+f(u))$ attains its maximum and minimum, respectively, with slope $s$.
\end{itemize}

\begin{figure}[!t]
\centering
\includegraphics[width=3.5in]{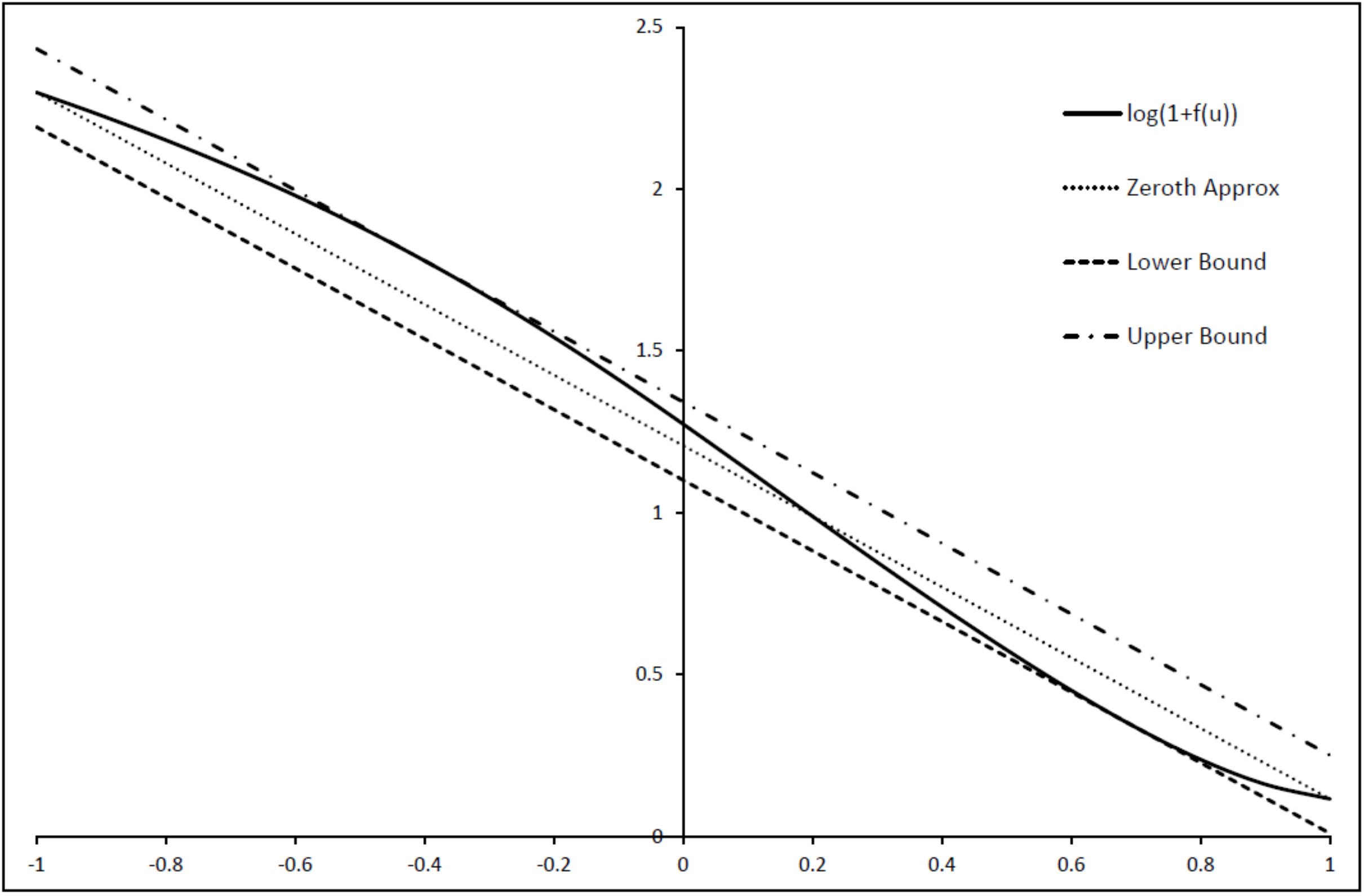}
\caption{Plot of $\log(1+f(u))$ in range $[-1,1]$ along with its zeroth-order straight line approximation and tangent upper and lower bounds.}
\label{fig:fooApprox}
\end{figure}

\subsection{Zeroth-order Approximation of the Optimal ANC Rate}
\label{subsec:zerothapprox}
The equation of the straight line passing through the points $[\log(1+f(-1)),-1]$ and $[\log(1+f(1)),1]$ is:
\begin{equation*}
y_z(u) = \halfrac\left[\log\frac{1+f(1)}{1+f(-1)}\right] u + \halfrac\log[(1+f(1))(1+f(-1))]
\end{equation*}
Substituting this in \eqref{eqn:infoRateFin2} and subsequently substituting the resultant expression in \eqref{eqn:maxAFrate}, we obtain the zeroth-order approximation of the optimal ANC rate as follows:
\begin{align}
I_{ANC}(P_s) &\approx \max_{\stackrel{\bm{\beta}:0 \le \beta_{m}^2 \le \beta_{m, max}^2}{1 \le m \le N}} \frac{1}{2 \pi} \int_{-1}^{1} \frac{1}{\sqrt{1-u^2}} y_z(u) du \label{eqn:zerothapprox} \\
             &= \max_{\stackrel{\bm{\beta}:0 \le \beta_{m}^2 \le \beta_{m, max}^2}{1 \le m \le N}} \frac{1}{4} \log[(1+f(1))(1+f(-1))] \nonumber
\end{align}

\subsection{First-order Approximation of the Optimal ANC Rate}
\label{subsec:firstapprox}
To compute the first-order approximations of the optimal ANC rate, we need to find two tangents bounding the curve $\log(1+f(u))$ in range $u \in [-1, 1]$ and passing through the points of maximum and minimum of the curve in the range, respectively, with their slopes equal to the slope s of the zeroth-order line. However, as there can be more than two such points where the slope of the curve is equal to the desired slope, we need to systematically search among all such points for the two desired values $u_{max}$ and $u_{min}$ of the integration variable $u$ at which $\log(1+f(u))$ attains its maximum and minimum, respectively with slope $s$. In the next page we provide an algorithm to compute these upper and lower bounds.

\textit{Remark 2:} In some instances, the zeroth-order straight line itself provides an upper or lower bound. For example, for a convex curve the zeroth-order line provides an upper bound, and for a concave curve it provides a lower bound.

\textit{Remark 3:} Apart from the zeroth and the first order straight-line approximations, the curve $\log(1+f(u))$ can be approximated with higher order polynomials. However, it leads, in general, to analytically intractable problems without substantially increasing the accuracy of the approximation.

In Figure~\ref{fig:fooApprox}, the zeroth and first-order approximations of $\log(1+f(u))$ in range $[-1,1]$ are illustrated with respect to one of its typical plots in Figure~\ref{fig:fooPlots}.

In the next section we discuss the computational complexity of these two approximation schemes for general nonlinear chains, starting with $(2,2)$ chains.

\newpage
\hspace{-1.0em}\hrulefill

\hspace{-0.5em}{\textbf{Algorithm:} First-order Approximation}

\vspace{-0.2cm}\hspace{-1.0em}\hrulefill
\begin{codebox}
\li Compute the slope $s$ of the zeroth-order line:
\zi $s = \halfrac\left[\log\frac{1+f(1)}{1+f(-1)}\right]$
\li Compute the roots of $\frac{d}{du} \log(1+f(u)) = s$.
\zi Let $R$ denote the set of the roots.
\li Compute $u_{max} = \argmax_R \log(1+f(u))$
\zi \hspace{0.5in} $u_{min} = \argmin_R \log(1+f(u))$
\li Equation of the tangent that bounds $\log(1+f(u))$ from
\zi above is: $y_{ub}(u) = s u + [\log(1+f(u_{max})) - s u_{max}]$
\li Equation of the tangent that bounds $\log(1+f(u))$ from
\zi below is: $y_{lb}(u) = s u + [\log(1+f(u_{min})) - s u_{min}]$
\li Compute the upper bound $UB$ on $I_{ANC}(P_s)$: substitute
\zi $y_{ub}(u)$ in \eqref{eqn:infoRateFin2} and subsequently substitute the resultant
\zi expression in \eqref{eqn:maxAFrate}:
\zi $UB = \max_{\bm{\beta}:0 \le \beta_{m}^2 \le \beta_{m, max}^2} \halfrac [\log(1+f(u_{max})) - s u_{max}]$
\li Compute the lower bound $LB$ on $I_{ANC}(P_s)$: substitute
\zi $y_{lb}(u)$ in \eqref{eqn:infoRateFin2} and subsequently substitute the resultant
\zi expression in \eqref{eqn:maxAFrate}:
\zi $LB = \max_{\bm{\beta}:0 \le \beta_{m}^2 \le \beta_{m, max}^2} \halfrac [\log(1+f(u_{min})) - s u_{min}]$
\End
\end{codebox}
\vspace{-0.2cm}\hrulefill

\section{Complexity of the Proposed Zeroth and First-order Approximation Schemes}
\label{sec:cmplxity_linearChains}
\subsection{$(2,2)$ Nonlinear Chain Networks}
\label{subsec:22_linchains}
Consider $(2,2)$ nonlinear chain network of Figure~\ref{fig:linChainExa}.

From the source, there are two paths with delay 1 to the destination, namely $\{(s1t),(s2t) \}$ and only one path $(s12t)$ with delay 2.

For the noise at the input to node 1, there is only one path $(1t)$ of delay 1 to the destination and another path $(12t)$ of delay 2. Similarly, for the noise at the input of node 2, there is only one path $(2t)$ of delay 1.

Therefore, we have the following expressions for the modified channel gains from the source and each relay node to the destination:
\begin{align*}
h_1 &=h_{s1}\beta_1h_{1t}+h_{s2}\beta_2h_{2t} \\
h_2 &=h_{s1}\beta_1h_{12}\beta_2h_{2t} \\
h_{1,1} &=\beta_1h_{1t} \\
h_{1,2} &=\beta_1h_{12}\beta_2h_{2t} \\
h_{2,1} &=\beta_2h_{2t}
\end{align*}

In terms of these modified channel gains the source-destination ISI channel in \eqref{eqn:ISIchnl} can be written as:
\begin{equation*}
y_t[n]=\sum\limits_{d=1}^2  h_d x_s[n-d] + \sum\limits_{d=1}^2  h_{1,d} z_1[n-d] + h_{2,1} z_2[n-1] + z_t[n],
\end{equation*}
where
\begin{align}
0 \le \beta_1^2 \le \beta_{1,max}^2 & = \frac{P_1}{h_{s1}^2P_s+\sigma^2} \label{eqn:22_feasibleRegion} \\
0 \le \beta_2^2 \le \beta_{2,max}^2 & = \frac{P_2}{(h_{s2}^2+h_{s1}^2\beta_1^2h_{s2}^2)P_s+(1+\beta_1^2h_{12}^2)\sigma^2} \nonumber
\end{align}

For this channel, for a given network-wide scaling factor $\bm(\beta)$, we have from \eqref{eqn:infoRateFin} and \eqref{eqn:trnsFn}:
\begin{equation*}
I(P_s,\bm{\beta}) = \frac{1}{2\pi} \int_0^{\pi} \log\Bigg[1+\frac{P_s}{\sigma^2} \frac{|H(\lambda)|^2}{1+\sum_{m=1}^2 |H_m(\lambda)|^2}\Bigg]d\lambda,
\end{equation*}
where
\begin{align*}
H(\lambda) &= \sum\limits_{d=1}^2 h_d e^{-id\lambda} = h_1 e^{-i\lambda} + h_2 e^{-2i\lambda}\\
H_1(\lambda) &=\sum\limits_{d=1}^2 h_{1,d} e^{-id\lambda} = h_{1,1} e^{-i\lambda} + h_{ 12} e^{-2i\lambda}\\
H_2(\lambda) &=\sum\limits_{d=1}^1 h_{2,d} e^{-id\lambda} = h_{2,1} e^{-i\lambda}
\end{align*}
Substituting these expressions for $H(\lambda), H_1(\lambda)$ and $H_2(\lambda)$ in the expression for $I(P_s,\bm{\beta})$ above, we have:
\begin{align*}
I(P_s,\bm{\beta}) &= \frac{1}{2\pi}\int_0^{\pi}\log\left(1+\frac{P_s}{\sigma^2}~\frac{ A_0+A_1\cos\lambda}{B_0+B_1\cos\lambda}\right)d\lambda \\
             &= \frac{1}{2\pi} \int_{-1}^{1} \frac{1}{\sqrt{1-u^2}} \log\left(1+\frac{P_s}{\sigma^2} \frac{A_0+A_1 u}{B_0+B_1 u}\right) du \\
             &= \frac{1}{2\pi} \int_{-1}^{1} \frac{1}{\sqrt{1-u^2}} \log\left(\frac{C_0+C_1u}{B_0+B_1u}\right) du
\end{align*}
where
\begin{align*}
A_0 &=h_1^2+h_2^2 \\
A_1 &=2 h_1 h_2 \\
B_0 &=1+h_{1,1}^2+h_{1,2}^2+h_{2,1}^2 \\
B_1 &=2 h_{1,1} h_{1,2} \\
C_0 &=B_0+\frac{P_s}{\sigma^2}A_0 \\
C_1 &=B_1+\frac{P_s}{\sigma^2}A_1
\end{align*}

\textit{Zeroth-order Approximation:} The equation of the zeroth-order straight line approximation of $\log\left(\frac{C_0+C_1u}{B_0+B_1u}\right)$ is:
\begin{equation*}
y_z(u) = s u + \halfrac\log\left(\frac{C_0^2 - C_1^2}{B_0^2-B_1^2}\right),
\end{equation*}
where
\begin{equation*}
s = \halfrac \log\left(\frac{C_0+C_1}{C_0-C_1}\frac{B_0-B_1}{B_0+B_1}\right)
\end{equation*}

Substituting this in the above expression for $I(P_s,\bm{\beta})$ and then further substituting the resultant expression in \eqref{eqn:maxAFrate}, we obtain the following zeroth-order approximation of the optimal ANC rate for $(2,2)$ chain:
\begin{equation}
\label{eqn:22_zeroth}
I_{ANC}(P_s) \approx \max_{\bm{\beta}:0 \le \beta_i^2 \le \beta^2_{i,max}} \frac{1}{4}\log\left(\frac{C_0^2 - C_1^2}{B_0^2-B_1^2}\right)
\end{equation}

Equating the first-order partial derivatives of the objective function with respect to $\beta_1$ and $\beta_2$, to zero, we obtain a system of two simultaneous polynomial equations. The stationary points of the objective function can thus be found by first computing a Gr\"{o}bner basis of the left hand-side of the equations to decide if this system of polynomial equations is zero-dimensional, in which case the solution can be obtained using various numerical or algebraic techniques \cite{102sturmfels}. Using second-order convexity condition \cite{106boydVandenberge}, we can determine for each such stationary point if it is the point of local minimum, local maximum, or a saddle point. We can prove that if the corresponding point of the global maximum of the objective function lies outside the feasible region in \eqref{eqn:22_feasibleRegion}, then the solution lies on the boundary of feasible region.

\textit{First-order Approximation:}
Let $u_{max}$ and $u_{min}$ be the values of the integration variable $u$ in the above expression for $I(P_s, \bm{\beta})$ such that the function $\log\left(\frac{C_0+C_1u}{B_0+B_1u}\right)$ attains its maximum and minimum values at it, respectively, with slope $s$. Then, the equation of the tangent that bounds $\log\left(\frac{C_0+C_1u}{B_0+B_1u}\right)$ from above with slope $s$ is:
\begin{equation*}
y_{ub}(u) = s u + \bigg[\log\left(\frac{C_0+C_1u_{max}}{B_0+B_1u_{max}}\right) - s u_{max}\bigg]
\end{equation*}
Similarly, the equation of the tangent that bounds $\log\left(\frac{C_0+C_1u}{B_0+B_1u}\right)$ from below with slope $s$ is:
\begin{equation*}
y_{lb}(u) = s u + \bigg[\log\left(\frac{C_0+C_1u_{min}}{B_0+B_1u_{min}}\right) - s u_{min}\bigg]
\end{equation*}
Substituting $y_{ub}(u)$ and $y_{lb}(u)$ in the above expression for $I(P_s,\bm{\beta})$ and then further substituting the resultant expression in \eqref{eqn:maxAFrate}, we obtain respectively the following upper and lower bounds the optimal ANC rate for $(2,2)$ chain:
\begin{align*}
UB &= \max_{\bm{\beta}:0 \le \beta_i^2 \le \beta^2_{i,max}} \halfrac\bigg[\log\left(\frac{C_0+C_1u_{max}}{B_0+B_1u_{max}}\right) - s u_{max}\bigg] \\
LB &= \max_{\bm{\beta}:0 \le \beta_i^2 \le \beta^2_{i,max}} \halfrac\bigg[\log\left(\frac{C_0+C_1u_{min}}{B_0+B_1u_{min}}\right) - s u_{min}\bigg]
\end{align*}

Following the procedure described above for computing the zeroth-order approximation, we can also compute the first-order upper and lower bounds on the optimum ANC rate for $(2,2)$ chain networks.

In the following section the optimal rate for $(2,2)$ nonlinear chain with varying source power $P_s$ is plotted along with its zeroth-order approximation and upper and lower bounds as derived above.

\subsection{$(N,k)$ Nonlinear Chain Networks}
\label{subsec:nk_linchains}
For general $(N,k)$ nonlinear chain networks, the procedure described in the last subsection for computing the zeroth and the first-order approximations is computationally inefficient because Gr\"{o}bner basis computation is ${\mathcal EXPSPACE}$-hard, in general \cite{113gathenGerhard}. However, the following lemma shows that for a class of general $(N,k)$ chain networks, the zeroth and first-order approximation can be computed efficiently.

\begin{pavikl}
\label{lemma:nk_zeroth}
Consider the $(N,k)$ nonlinear chain where the gains along all outgoing channels for each node $i, i \in \{s\} \bigcup R$, are equal to $h_i$. For such a network, the zeroth order approximation to the optimal ANC rate can be computed in polynomial time. In other words, for such a network, we have the following equivalence
\begin{equation*}
\max_{\bm{\beta}} \frac{1}{4}\log\left(\frac{C_0^2 - C_1^2}{B_0^2-B_1^2}\right) = \max_{\beta_1}\ldots\max_{\beta_N} \frac{1}{4}\log\left(\frac{C_0^2 - C_1^2}{B_0^2-B_1^2}\right)
\end{equation*}
\end{pavikl}
\begin{IEEEproof}[Proof (Sketch)]
It can be proved that for a given $(\beta_1, \ldots, \beta_{N-1})$, the objective function is convex in $\beta_N$. In general, for a given $(\beta_1, \ldots, \beta_{i-1})$, the objective function is convex in $\beta_i, i > 1$. This allow us to reduce a joint optimization problem over $N$-dimensional vectors $\bm{\beta}: (\beta_1, \ldots, \beta_N)$ to $N$ successive convex optimization problems.
\end{IEEEproof}

\section{Performance Analysis}
\label{sec:perfAnalysis}
In this section, we evaluate the performance of both, the zeroth and first order approximation to the optimal ANC rate for various nonlinear chain networks using the procedure described in the last section.

\begin{figure}[!t]
\centering
\includegraphics[width=3.45in]{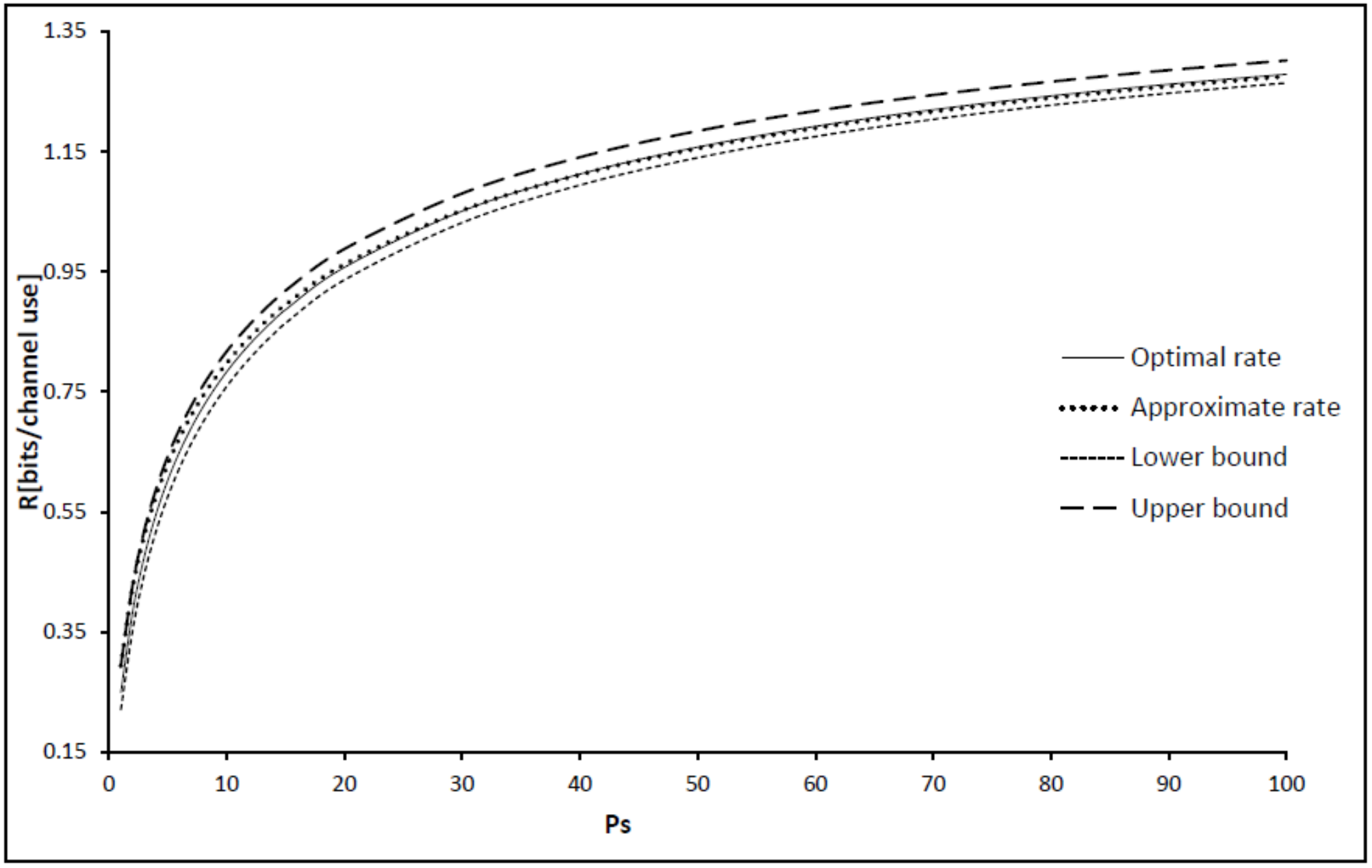}
\caption{Plot of the optimal rate for $(2,2)$ nonlinear chain with varying source power $P_s$ along with its zeroth-order straight line approximation and first-order tangent upper and lower bounds. Each data point is averaged over $100$ network instances.}
\label{fig:22Plots}
\end{figure}

In Figure~\ref{fig:22Plots} the optimal rate for $(2,2)$ nonlinear chain with varying source power $P_s$ is plotted along with its zeroth-order approximation and upper and lower bounds as derived Section~\ref{sec:cmplxity_linearChains}. It can be observed that the zeroth-order approximation in this case is very accurate and so are the upper and lower bounds which have a gap no more than $0.04$ bits/channel use.

\begin{figure}[!t]
\centering
\includegraphics[width=3.45in]{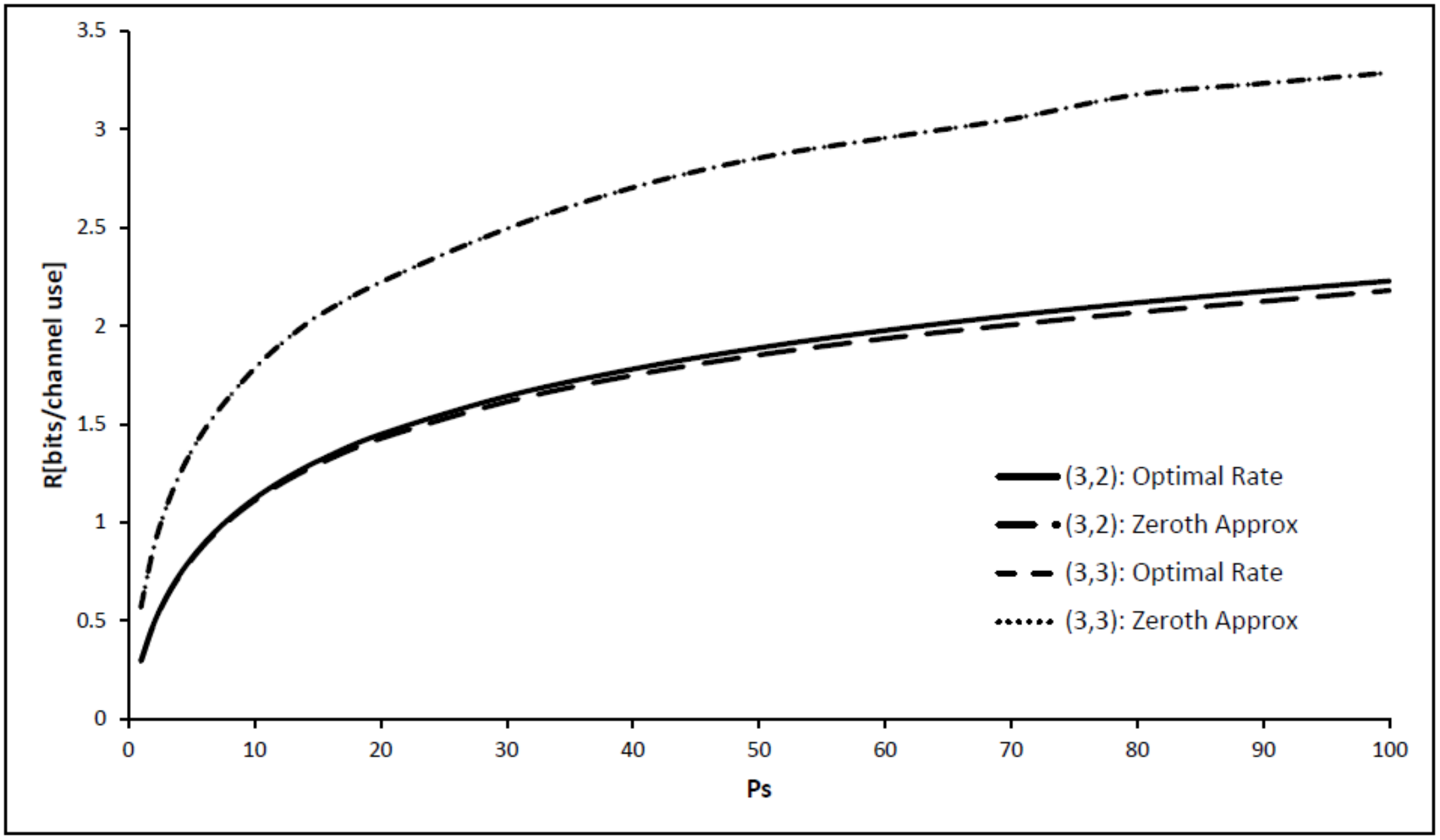}
\caption{Plot of the optimal rate for $(3,2)$ and $(3,3)$ nonlinear chains with varying source power $P_s$ along with their zeroth-order straight line approximations.}
\label{fig:3zeroth}
\end{figure}

In Figure~\ref{fig:3zeroth} the optimal rates for $(3,2)$ and $(3,3)$ nonlinear chains are plotted versus the source power $P_s$ along with the corresponding zeroth-order straight line approximations. It can be observed that zeroth-order approximation tightly approximates the corresponding optimal rates within $0.05$ bits/channel use.

\begin{figure}[!t]
\centering
\includegraphics[width=3.45in]{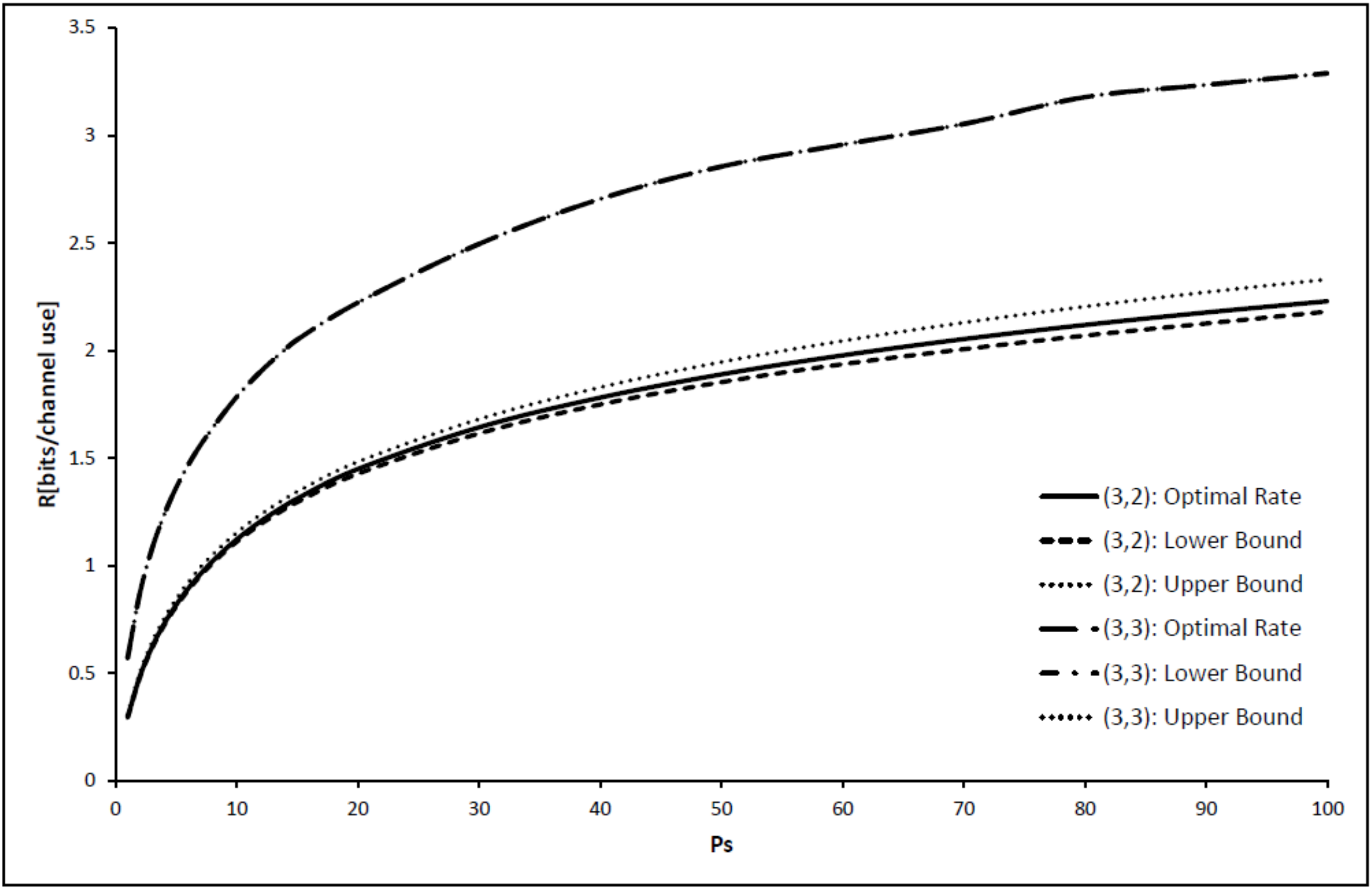}
\caption{Plot of the optimal rate for $(3,2)$ and $(3,3)$ nonlinear chains with varying source power $P_s$ along with their first-order tangent upper and lower bounds.}
\label{fig:3first}
\end{figure}

In Figure~\ref{fig:3first} the optimal rates for $(3,2)$ and $(3,3)$ nonlinear chains with varying source power $P_s$ are plotted along with their corresponding upper and lower bounds obtained from the first-order approximations. It can be observed that the upper and lower bounds have a gap no more than $0.15$ bits/channel use.

\section{Conclusion and Future Work}
\label{sec:conclFW}
The problem of characterizing the optimal rate achievable with analog network coding (ANC) for a unicast communication over general wireless relay networks is computationally hard. To gain a better understanding of the problem to construct low-complexity schemes to characterize the optimal ANC rate for a much wider class of general wireless relay networks, in this paper we proposed a twofold approach of nonlinear chain networks and two approximation schemes. This approach leads to polynomial-time tight characterization of the optimal ANC rate for a few classes of non-layered networks which could not be so addressed using existing schemes. In the future, we plan to construct analytical characterizations for the performance of the two proposed schemes and extend those to general wireless relay networks.

\end{document}